\documentclass[10pt,letterpaper]{article}
\usepackage[top=0.85in,left=2.75in,footskip=0.75in]{geometry}

\usepackage{amsmath,amssymb}

\usepackage{changepage}

\usepackage{textcomp,marvosym}

\usepackage{cite}

\usepackage{nameref,hyperref}

\usepackage[nopatch=eqnum]{microtype}
\DisableLigatures[f]{encoding = *, family = * }

\usepackage[table]{xcolor}

\usepackage{array}

\usepackage{color,soul}

\newcolumntype{+}{!{\vrule width 2pt}}

\newlength\savedwidth

\usepackage[aboveskip=1pt,labelfont=bf,labelsep=period,justification=raggedright,singlelinecheck=off]{caption}

\makeatletter
\renewcommand{\@biblabel}[1]{\quad#1.}
\makeatother

\usepackage{lastpage,fancyhdr,graphicx}
\usepackage{epstopdf}
\fancyheadoffset[L]{2.25in}
\fancyfootoffset[L]{2.25in}
\begin{document}
\vspace*{0.2in}

\begin{flushleft}
{\Large
\textbf\newline{Inference of dynamical gene regulatory networks from single-cell data with physics informed neural networks} 
}
\newline
\\
Maria Mircea\textsuperscript{1,\textcurrency a},
Diego Garlaschelli\textsuperscript{2,3},
Stefan Semrau\textsuperscript{1,\textcurrency b, *}
\\
\bigskip
\textbf{1} Department of Physics, Leiden University, Leiden, The Netherlands
\textbf{2} Lorentz Institute, Leiden University, Leiden, The Netherlands
\textbf{3} IMT, Lucca, Italy
\bigskip

\textcurrency a Current Address: Life and Medical Sciences Institute, University of Bonn, Bonn, Germany \\
\textcurrency b Current Address: New York Stem Cell Foundation Research Institute, New York, USA

* semrau@physics.leidenuniv.nl

\end{flushleft}
\section*{Abstract}
One of the main goals of developmental biology is to reveal the gene regulatory networks (GRNs) underlying the robust differentiation of multipotent progenitors into precisely specified cell types. Most existing methods to infer GRNs from experimental data have limited predictive power as the inferred GRNs merely reflect gene expression similarity or correlation. Here, we demonstrate, how physics-informed neural networks (PINNs) can be used to infer the parameters of predictive, dynamical GRNs that provide mechanistic understanding of biological processes. Specifically we study GRNs that exhibit bifurcation behavior and can therefore model cell differentiation. We show that PINNs outperform regular feed-forward neural networks on the parameter inference task and analyze two relevant experimental scenarios: 1. a system with cell communication for which gene expression trajectories are available and 2. snapshot measurements of a cell population in which cell communication is absent. Our analysis will inform the design of future experiments to be analyzed with PINNs and provides a starting point to explore this powerful class of neural network models further.



\section{Introduction}
\begin{figure}[!h] 
    \centering
    \includegraphics[width=1\textwidth]{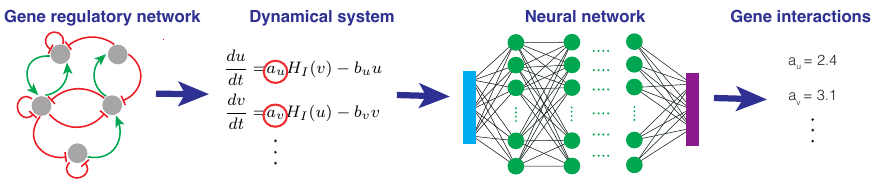}
    \caption{{\bf Gene regulatory network (GRN) inference with neural networks (NNs)} First, a particular topology of the GRN is assumed. Together with the functional form of the interactions, the GRN topology defines a set of differential equations with undetermined parameter values. Next, a NN is trained on experimental or simulated time-series data. The parameters learned during training set the strength of interactions between genes. The fully determined dynamical system can then be used for predictions.}
    \label{figures-intro}
\end{figure}

Since the advent of single-cell molecular profiling, developmental biology has been inundated with high-dimensional data we are still learning to make sense of. Various machine learning methods have been used to find patterns in single-cell data, such as cell types or differentiation paths \cite{Yu2022, Saelens2019}. Notwithstanding the great success of these methods, it remains difficult to infer mechanistic insights or quantitative, predictive models from single-cell data. Yet, one of the main goals of developmental biology is to understand the gene regulatory mechanisms underlying the robust differentiation of precisely defined cell types from multipotent progenitors \cite{Mircea2021}.\\
A common approach to the predictive mathematical modeling of differentiation uses the framework of dynamical systems theory \cite{Guillemin2020, Greulich2020, Xu2020, Huang2012}. In the context of differentiation, the dynamical system governs the abundance of gene products in the cell and stable attractor states are interpreted as cell types. Under certain conditions, the system can be represented by a quasi-potential \cite{Zhou2012}. This potential is the mathematical equivalent of Waddington's landscape \cite{Waddington1975}, a seminal, qualitative model of differentiation in which the valleys in the landscape correspond to different cell types. In most models, the dynamical system has the structure of a network, in which the nodes are gene products, typically transcription factors, and the edges indicate interactions between them. Ideally, such a gene regulatory network (GRN) model should be able to predict the outcome of a differentiation process, given the initial cell state and external cues. Simulations using small GRNs with $2-5$ nodes indeed exhibit bifurcations resembling actual differentiation processes  \cite{Ferrell2012,Saiz2020,Raina2020,Franke2021, Huang2007}. \\
As the parameter space grows quickly with the the size of a GRN, it can be tedious to find regimes with relevant behavior. A large body of work has therefore been devoted to inferring GRNs from measurements, typically transcriptomics or proteomics data sets. Most recently, single-cell data has been leveraged to that end \cite{Pratapa2020}. Many inference methods use measures of similarity or correlation between genes and prior biological knowledge, most often about protein-protein binding affinities or the targets of transcription factors \cite{Huynh-Thu2010,Aibar2017,Papili2018,Qiu2020}. These methods can infer the existence of correlative or even causal relationships, especially if chromatin accessibility is taken into consideration \cite{Kamimoto2020}. However, they are typically unable to infer interaction strengths and are thereby lacking in predictive power. In fact, if only single-cell snapshot data is used and there is no prior biological knowledge, there are fundamental limits to GRN inference \cite{Weinreb2018}. One should therefore 1. use time-resolved data that ideally contains information about the trajectories of individual cells and 2. constrain the inference problem with assumptions about the GRN. Seminal work using a Boolean network approach \cite{Dunn2014} or, more recently, catastrophe theory and approximate Bayesian computation \cite{Saez2021}, have successfully inferred predictive GRN models from time resolved data. \\
Another class of machine learning tools that have become extremely important in many fields are neural networks (NNs). These have been highly successful in pattern recognition and classification tasks \cite{Abiodun2019} and are used extensively to interpret single-cell omics data \cite{Angermueller2016NN,Eraslan2019}. Naturally, NNs have also been used to infer GRNs from measurements \cite{Yang2019, Wang2020b}. However, existing NN methods require GRNs obtained by other means as training data, which might limit the fidelity of the inferred GRNs. The optimal NN method would only use gene expression as training data while allowing us to implement prior knowledge or assumptions about the GRN to make the inference problem feasible. The recently developed physics-informed neural networks (PINNs) \cite{Raissi2019,Lu2019,Karniadakis2021} enable us to do just that. PINNs can solve a broad range of differential equations and also infer undetermined paramters. They have been applied successfully to various systems biology tasks \cite{Lagergren2020, Yazdani2020, AlQuraishi2021}.  \\
In this study, we explore in how far PINNs can be used to infer GRNs. In our case, the differential equations to be solved by the PINN are defined by GRN topology and the mathematical expressions describing the interactions between the genes. Given gene expression measurements as training data, PINNs should be able to infer gene interactions. Fig. \ref{figures-intro} shows a schematic of the inference procedure. Once the parameters have been learned, the dynamical system can then be used to make predictions. PINNs should thus allow us to gain mechanistic insights from measured expression data. As a proof of concept, we simulate data based on a GRN model recently introduced by Stanoev et al. \cite{Stanoev2021}. At its core this GRN has two mutually inhibiting genes, $u$ and $v$, which can be seen as the master transcription factors governing two alternative cell fates. This network motif has been studied intensively since it is one of the simplest motifs that exhibit bifurcations \cite{Ferrell2012,Bessonnard2014, Huang2007, Huang2012}: Depending on the network parameters, a single stable attractor, interpreted as a multilineage primed (mlp) state can split into two stable attractors, which correspond to two different lineages. Such bifurcations successively create the large diversity of cell types in an adult organism, starting from the one-cell embryo \cite{Xia2019}. In addition  to the cell intrinsic mutual inhibition of the master transcription factors $u$ and $v$, the model by Stanoev et al. \cite{Stanoev2021} also implements cell communication, which plays an essential role in development \cite{Giacomelli2020,Berenger-Currias2022}. In the Stanoev model, cells communicate via a diffusible signaling molecule $s$. This molecule is assumed to be activated by $u$ and in turn inhibits $u$ in both an autocrine and paracrine manner. The differential equations defining this GRN are shown in Fig. \ref{figures-1}a. With this system, Stanoev et al. showed that population size can effectively serve as a control parameter that can bring the system from a homogeneous, progenitor state to a heterogeneous, differentiated state \cite{Stanoev2021}. This mechanism is an interesting alternative to the previously suggested noise-driven fate decisions \cite{Elowitz2002, Safdari2020}. In certain parameter regimes, the GRN also creates regular, Turing-like spatial patterns. \\

In this study, we first explore the qualitative behavior of the GRN introduced by Stanoev et al. \cite{Stanoev2021}. Subsequently we demonstrate that a naive feed-forward NN architecture and training based on simulations in a limited parameter regime are insufficient for robust GRN inference. We then explore how accurately PINNs can infer GRN parameters. Surprisingly, we find that it is not necessary to use all variables of the GRN for training. In other words, the measurement of a subset of genes across time can be sufficient for GRN inference. Lastly, we investigate a simpler system without cell communication. We study in how far GRN inference is still possible, if only snapshot data at discrete time points is available. This scenario is highly relevant as it describes the typical single-cell profiling experiment. This manuscript thus provides a thorough assessment of PINNs for the purpose of GRN inference.

\section{Materials and methods}
Python (V 3.9) was used for all computations. NNs were implemented with \textit{tensorflow} (V.2.6). All figures were generated with R (V 4.1). 

\subsection{Inference of a GRN with cell communication from trajectories}
\subsubsection{Differential equations}
The set of coupled differential equations in Fig. \ref{figures-1}a were adopted from \cite{Stanoev2021}. For each cell, there are two mutually inhibiting genes, $u$ and $v$. Additionally, a signalling molecule $s$ that is stimulated by $u$, inhibits $u$ in an autocrine and paracrine way. In the model considered here, all signalling molecules of neighboring cells contribute equally. The dynamics of each cell $i \in \{1, ..., N\}$ is governed by this system of differential equations
\begin{align}
\begin{split}
\frac{du_i}{dt} &= a_u H_I(v_i) + a_{us} H_I(s_{ext}) - u_i \\ 
\frac{dv_i}{dt} &= a_v H_I(u_i)  - v_i \\
\frac{ds_i}{dt} &= a_s H_A(u_i) - s_i, 
\end{split}
\end{align}
where $s_{ext}$ denotes signalling molecule abundance averaged over the neighbors: $s_{ext} = \frac{1}{k+1}\sum_{j \in N(i) \cup i} s_j$ with $k$ the number of cells in the neighbourhood $N(i)$ of cell $i$. $H_I$ and $H_A$ are the inhibiting and activating Hill functions, respectively, defined here with a fixed Hill coefficient of $2$:
\begin{align}
H_I(x) = \frac{1}{1 +x^2}, ~~
H_A(x) = \frac{x^2}{1 + x^2}
\end{align}
This results in a set of coupled differential equations with $3N$ equations and $3N$ variables. The parameters $a_u, a_v, a_s$ and $a_{us}$ are the same for each cell. Time was rescaled by an inverse degradation rate which was assumed to be identical for all genes. An additional parameter $\lambda$, used by Stanoev et al. \cite{Stanoev2021} to control the speed of the temporal evolution, was set to $1$.\\
We distributed the cells on a regular grid, generated with \textit{python-igraph} (V 0.9), and cell communication was typically restricted to nearest neighbors, unless otherwise indicated by edges in the graph. 
\subsubsection{Steady states}
Steady states were found with a multi-start optimization algorithm using \textit{scipy} (V 1.7). The stability was calculated based on the Jacobian matrix evaluated around the steady state. Bifurcation analysis was performed by repeating the optimization algorithm for each value of the control parameter (either $a_u$ or $a_{us}$). For the 2-cell and 4-cell configurations, 500 initial values were chosen uniformly in the interval $ [0, 3] $ for each of the $3N$ variables. In the study of the effect of cell number (Fig. \ref{figures-1}e,f) $100$ initial values were used. \\
Steady states were either mlp, where each cell was an mlp, or differentiated, consisting of u-high and v-high cells. The mlp steady state was identified based on comparison with gene expression in the 1-cell configuration with the same parameters. If the relative error (comparing to the 1-cell mlp state) for all dependent variables was below $0.1$, a steady state was annotated as mlp. To identify u-high and v-high cells in a differentiated steady state, the ratio of $u$ and $v$ in individual cells was compared to the ratio of $u$ and $v$ in the mlp steady state for the same parameters. If the ratio of $u$ and $v$ in a cell was larger than in the mlp steady state, the cell was consider to be a 'u-high' cell, if the ratio was smaller than in the mlp steady state, the cell was considered to be 'v-high'.   

\subsubsection{Data simulation} \label{dat_sim}
Data points were generated by numerical integration (NI) of the differential equations for a given set of parameters. We used \textit{scipy} (V 1.7) with the explicit Runge-Kutta method for integration. The initial conditions were chosen randomly from a uniform distribution in the interval $[0,1]$. Numerical integration was performed on $100$ equidistant time points in the interval $[0, T]$. When the entire trajectory was used for training, a subset of $25$ equidistant time points was used. When only the initial and final state were used as input, we considered the values at $t = 0$ and $t=T$. 

\subsubsection{Feed-forward neural network}
The feed-forward NN architecture is depicted in Fig. \ref{figures-2}a. The NN input takes 25 time points for each variable. The GRN consisted of 4 cells, which all communicate with each other, resulting in 12 independent variables in total. Thus, the input layer of the NN consists of 300 nodes. We found that a 20\% dropout rate in the input layer prevented over-fitting during training. The NN has 4 fully-connected hidden layers with 32 nodes each and Rectified Linear Unit (ReLu) activation functions. The output layer has 4 nodes for the results shown in  Fig. \ref{figures-2}d, corresponding to the parameters $a_u$,$a_v$,$a_s$ and $a_{us}$, and one node for the result shown in Fig. \ref{figures-2}e, corresponding to the parameter $a_u$. The loss function was defined using the mean absolute error, which slightly outperformed the mean squared error. For optimization the Adam optimizer was used. \\
For training and testing, trajectories were generated by numerical integration using randomly drawn initial states from the interval $[0,1]$. For Fig. \ref{figures-2}d, 1000 training trajectories were generated with parameters chosen randomly from uniform distributions over the following intervals: $a_u \in [1,3]$, $a_v \in [2, 5]$, $a_s \in [1, 3]$ and $a_{us} \in [0,2]$. To generate trajectories for testing in Fig. \ref{figures-2}d, parameters were chosen from the same parameter ranges. For each run, 3 out of the 4 parameters were kept fixed ($a_u = 2.4$, $a_v = 3.5, a_u = 2, a_{us} = 1$), while the remaining parameter was varied. For each parameter, 50 values were taken from the above intervals equidistantly and each set of parameters was used 20 times with different initial values for the numerical integration.\\
For Fig. \ref{figures-2}e, 1000 training trajectories were generated with three parameters fixed ($a_v = 3.5, a_u = 2, a_{us} = 1$) and $a_u$ sampled from the interval $[2.4, 2.7]$, where the dynamical system has multiple steady states in the 4-cell configuration. The trajectories used for testing were generated with the same fixed parameters, but 50 values of $a_u$ were equidistantly drawn from the interval $[1,3]$. Each parameter value was used 20 times with different initial conditions for the numerical integration.

\subsubsection{Physics Informed Neural Network}
All PINNs were implemented using \textit{DeepXDE} (V 0.14) \cite{Lu2019}. The network architecture is shown in Fig. \ref{figures-3}a. The input layer consists of only one node, which corresponds to the time $t$. The hidden layers are 4 fully connected layers with 40 nodes each. The output layer has $3N$ nodes, where $N$ is the number of cells, corresponding to the $3N$ dependent variables of the dynamical system. In Fig. \ref{figures-4}, a configuration of 4 cells that all communicate with each other was implemented, which results in an output layer of size $12$. \textit{tanh} was used as the activation function.\\
The loss function consists of three terms. The first penalizes deviation from the differential equations: 
\begin{align}
\begin{split}
\biggl(\frac{du_1}{dt} - a_u H_I(v_1) + a_{us} H_I(s^1_{ext}) - u_1\biggr)^2 &+ 
\biggl(\frac{dv_1}{dt} - a_v H_I(u_1)  - v_1\biggr)^2  \\  
 &+ \biggl(\frac{ds_1}{dt} - a_s H_A(u_1) - s_1\biggr)^2 +\\
\hdots \\
\biggl(\frac{du_n}{dt} - a_u H_I(v_n) + a_{us} H_I(s^n_{ext}) - u_n\biggr)^2 &+ 
\biggl(\frac{dv_n}{dt} - a_v H_I(u_n)  - v_n\biggr)^2 \\
&+ \biggl(\frac{ds_n}{dt} - a_s H_A(u_n) - s_n\biggr)^2
\end{split}
\end{align}
The second loss term considers the initial conditions, using the mean squared error. The last term in the loss function includes the training data, also using the mean squared error. For time points within the trajectory we defined boundary conditions using the \textit{PointSet} object from the \textit{DeepXDE} package. This object allows the user to supply measured data at any point in the input domain and the coresponding loss term will be added to the loss function. In order to specify the final time point in the \textit{DeepXDE} framework, we used the Dirichlet boundary condition object. In this way, we could set values for the time domain at the initial point $t=0$ and the final point $t=T$. \\
To create the results shown in Fig. \ref{figures-4}, PINNs were trained using $84$ different scenarios, $10$ times each, with randomly selected initial conditions for the differential equations. Results were averaged over the $10$ simulations. The following properties of the training data and PINN were varied:\\
\begin{itemize}
\item[(a)] number of time points used for training: 25 (full trajectory) or 2 (initial and final state)
\item[(b)] noise level: no noise; Gaussian with mean 0, standard deviation 0.1; Gaussian with mean 0, standard deviation 0.2. Negative values resulting from addition of noise addition were set to 0.
\item[(c)] number of dependent variables used for training: 1,2 or 3 
\item[(d)] identity of dependent variables used for training: [u,v,s], [u,v], [u,s], [v,s], u, v, s
\item[(d)] weights: no weights or ODE loss weighted with factor 1000
\end{itemize}

\subsubsection{Neural network validation}
We used three measures to quantify the performance of the PINN. First, we computed the relative error between the inferred parameters and the true parameters. Second, we calculated the mean squared error between the PINN approximation of the trajectories and trajectories obtained by  numerical integration using parameters and initial conditions inferred by the PINN. Lastly, we considered the test loss. 

\subsection{Inference of a GRN without cell communication from snapshot data}
\subsubsection{Differential equations}
Two mutually inhibiting genes, $u$ and $v$, were modeled with expressions used to describe lateral inhibition \cite{Ferrell2012}. The differential equations are given by: 
\begin{align}
\begin{split}
\frac{du}{dt} &= a_u \frac{1}{1 +( I_u v)^3} - b_u u = f_1(u,v)\\
\frac{dv}{dt} &= a_v \frac{1}{1 +( I_v u)^3} - b_v v = f_2(u,v)
\end{split}
\label{eq-ode-nocomm}
\end{align}
We used a set of parameters for which this dynamical system is bistable: $a_u = a_v = 1.5$, $I_u = I_v = 0.5$ and $b_u = b_v = 0.5$. To model inhibition we used an inhibiting Hill function with Hill coefficient $3$. \\
We describe the system at the population level with the joint probability density for the abundance of $u$ and $v$. The time evolution of this probability density is governed by the conservation of probability:
\begin{align}
\frac{\partial p}{\partial t} + \nabla p \cdot f(u,v) + p ~ div(f(u,v)) = 0,
\label{pde}
\end{align}
where $f(u,v)$ is defined as 
\begin{align}
f(u,v) = \begin{bmatrix}
f_1(u,v) \\
f_2(u,v)
\end{bmatrix}
\end{align}
Plugging in $f(u,v)$ gives the following differential equations:
\begin{align}
\frac{dp}{dt} + \frac{dp}{du} \left(a_u \frac{1}{1 +( I_u v)^4} - b_u u \right) + \frac{dp}{dv} \left(a_v \frac{1}{1 +( I_v u)^4} - b_v v\right) - (b_u + b_v) p = 0
\end{align}

\subsubsection{Data simulation}
Training data was created by numerical integration of the ODE in equation \ref{eq-ode-nocomm} as described in section \ref{dat_sim}. The initial conditions were sampled from a normal distribution with mean $1.5$ and standard deviation $0.1$. Trajectories with $50$ time points in the interval $[0, 20]$ were created. $4$, $8$ or $20$ equidistant time points were used for further computations. In order to create a probability density function, $1000$ trajectories were generated. The probability density was then estimated by the relative frequencies calculated for either $10$ bins or $20$ bins for each time point. The values for $u$ and $v$ in each bin were taken as the bin's midpoint. 

\subsubsection{Physics Informed Neural Network}
A PINN with the architecture shown in Fig. \ref{figures-5}a was implemented with \textit{DeepXDE}. The PINN takes three independent variables as input ($u$, $v$ and $t$). The geometry of the input space consisted of a rectangle with values between $0.5$ and $3.5$ on each side, and a time domain between $0$ and $20$. The hidden layers are $3$ fully connected layers with $32$ nodes each and \textit{tanh} activation functions. The output layer has only one node, which corresponds to the probability density $p(u,v,t)$. We applied an absolute value transformation to the output layer to ensure positive results for the probability density. Initial conditions were defined at $t = 0$ as a bivariate normal distribution with mean $ \begin{bmatrix} 1.5 \\ 1.5 \end{bmatrix}$ and variance $\begin{bmatrix} 0.1 & 0\\ 0 & 0.1 \end{bmatrix}$. Simulated training data was added, as described previously, with the \textit{PointSet} object in \textit{DeepXDE}. As before, the loss function was composed of 3 terms that consider the differential equation given in equation \ref{pde}, the initial conditions and the training data, using the mean squared error. Additionally, we implemented a soft constraint to enforce the integral of the probability density over the space domain to be equal to $1$ to ensure that the PINN produces a proper  probability density. To that end, we defined one operator boundary condition considering all values of the probability density inside the rectangular geometry, separately for each time point. The loss function was then defined as the mean squared error between the mean over all values of the probability density and the value $1$. This sampling strategy corresponds to a Monte Carlo simulation of the integral over the space domain which we equate to $1$. \\
For training, $500$ points were chosen randomly from the joint domain of $u$ and $v$ to define the initial condition loss, $2500$ points from the joint domain of $u$, $v$ and $t$ were chosen randomly to define the differential equation loss, and $400$ points were chose to fulfill the boundary condition loss. \\
The following properties of the training data and PINN were varied:
\begin{itemize}
    \item[(a)] Number of time points for the training data: 4, 8, or 20 equidistant time points
    \item[(b)] Noise level: no noise; Gaussian with mean 0, standard deviation 0.1; Gaussian with mean 0, standard deviation 0.2. Gaussian noise was added to the trajectories resulting from equation \ref{eq-ode-nocomm}.
    \item[(c)] Number of bins for relative frequencies: 10 or 20
    \item[(d)] We tested $4$ different weighting strategies, for the $4$ loss functions: PDE loss, IC loss, data loss and normalisation loss. The weights are given in the same sequence for the losses: (1) 1-1-1-1, (2) 1-1-1000-1, (3) 1000-1000-1-1000 and (4) 1-1000-1-1000.
\end{itemize}
 The parameters $a_u$, $a_v$, $b_u$ and $b_v$ were fixed and the parameters $I_u$ and $I_v$ were inferred by the PINN. 

\section{Results}
\subsection{Cell communication drives bifurcations in a GRN model of differentiation}
\begin{figure}[!h] 
    \centering
    \includegraphics[width=1\textwidth]{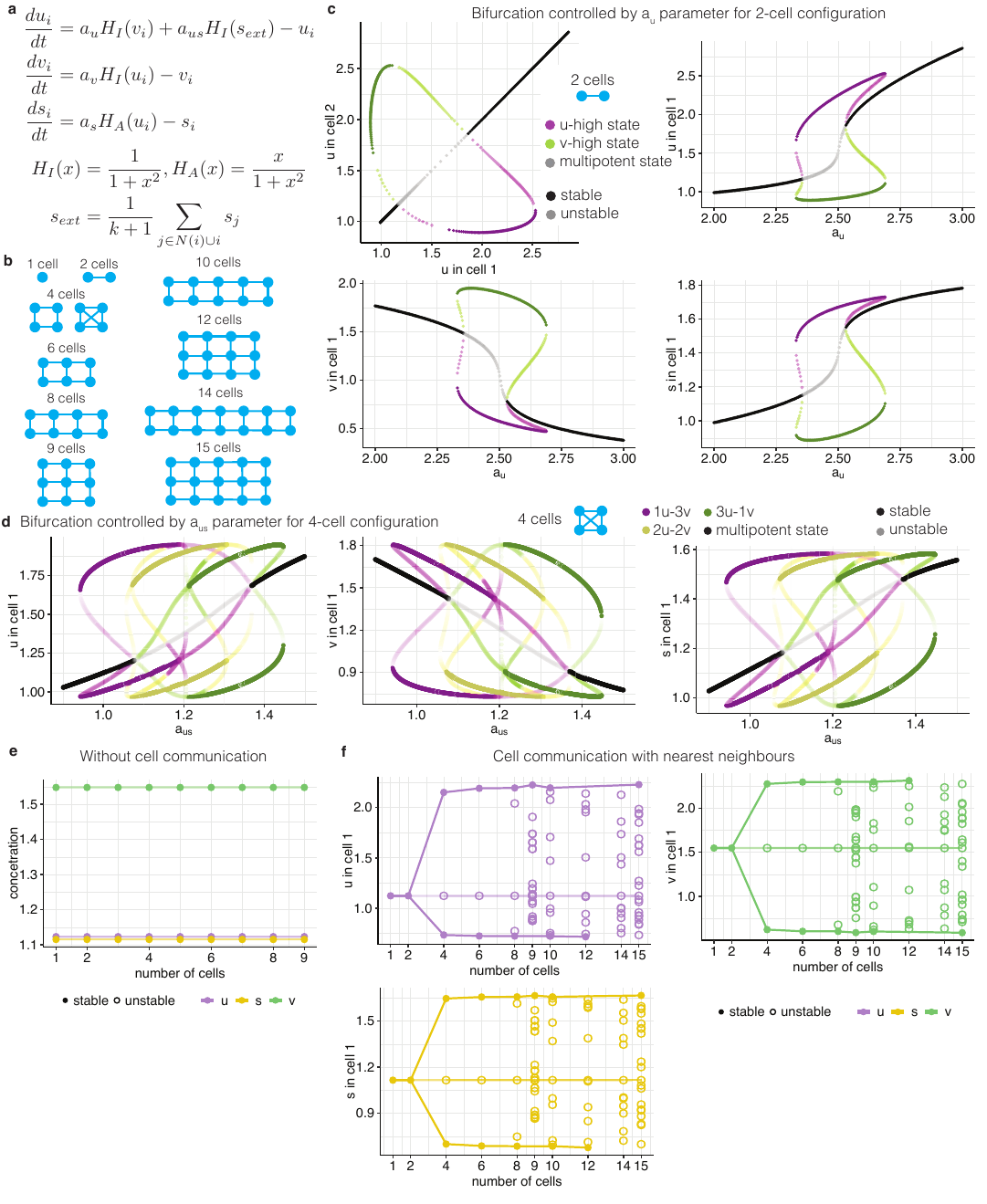}
    \caption{{\bf Cell communication drives bifurcations in a GRN model of differentiation.} See next page for caption.}
    \label{figures-1}
\end{figure}

We first set out to recapitulate the qualitative behavior of the GRN reported by Stanoev et al. \cite{Stanoev2021} to find interesting parameter regimes and establish a ground truth for GRN inference. We placed between 1 and 15 cells on a regular square grid and allowed cell communication between  nearest neighbors, unless otherwise indicated by edges in \ref{figures-1}b). The corresponding system of differential equations (shown in Fig. \ref{figures-1}a) was solved by numerical integration.\\

\addtocounter{figure}{-1}
\begin{figure}[!h] 
    \centering
    \caption{{\bf Cell communication drives bifurcations in a GRN model of differentiation.} a: System of differential equations corresponding to the GRN model by Stanoev et al. \cite{Stanoev2021} The mutual inhibition of the two master transcription factors $u$ and $v$ as well as the inhibition of $u$ by the signaling molecule $s$ are modeled with repressive Hill functions $H_I$. The cell autonomous activation of signalling molecule $s$ by $u$ is modeled with an activating Hill function $H_A$. $i$ is the cell index. $s_{ext}$ is the level of $s$ averaged over cell $i$ and its neighbors (typically nearest neighbors, unless otherwise indicated by the edges in panel b). The degradation rate for $u$, $v$ and $s$ is assumed to be identical, and time was rescaled with the inverse degradation rate, so that the rate does not appear explicitly in the equations. b: Studied configurations of cells. Edges indicate cell communication. c: Results for the 2-cell configuration. Several bifurcations are driven by the parameter $a_u$, which sets the strength of the inhibition of $u$ by $v$. d: Results for the 4-cell configuration with communication between all cells. Bifurcations are controlled by the parameter $a_{us}$ which determines the strength of inter-cellular communication. Colors distinguish stable states with different ratios of $u$- and $v$-high cells. e,f: Steady states (both stable and unstable) for the cell  configurations shown in panel b without cell communication (panel e) or with cell communication (panel f). The following parameters were used: $a_u = 2.4, ~ a_v = 3.5, ~ a_s = 2, a_{us} = 1$.}
\end{figure}
\setlength\parindent{0pt}
In the two-cell configuration, the parameter $a_u$, which sets the inhibition of $u$ by $v$ is a control parameter that can elicit a bifurcation (Fig. \ref{figures-1}c). For low values of $a_u$ there is one stable state. In this state, both cells have identical concentrations of $u$ and $v$. Following Stanoev et al. \cite{Stanoev2021}, we interpret a homogeneous state, where cells have identical, intermediate expression of both $u$ and $v$ as the mlp state. At a particular, critical value of $a_u$, two additional stable states appear through saddle node bifurcations. In one of these states, cell 1 has a high level of $u$ but a low level of $v$, while cell 2 has a low level of $u$ but a high level of $v$. In the other stable state, the expression patterns of cell 1 and 2 are reversed. These two states are thus considered differentiated states. Due to the mutual inhibition between $u$ and $v$ we will always find anti-correlation between the two genes, outside of the mlp state. Increasing $a_u$ further, at a second critical point, the mlp state becomes unstable through a subcritical pitchfork bifurcation. Between $a_u$ values of $2.33$ and $2.69$ the two differentiated states are the only stable states of the system. In summary, when controlled by $a_u$ the system goes from the mlp state for low values of $a_u$ via a small interval with three stable states (mlp, two differentiated states) to a wider interval with the two differentiated states as the only stable states. In other words, for differentiation to occur, a certain level of mutual inhibition between $u$ and $v$ is necessary. At even higher values of $a_u$ additional bifurcations occur and the system returns to a single stable state. We will not explore this behavior at high $a_u$ any further here, since it is unphysiological: In the real biological system, differentiated states are likely stabilized by other means (such as epigenetic marks), which would prevent the reversal to a single stable state.   
For larger numbers of cells, the behavior of the system becomes more complex. We studied in detail a population of 4 cells where each cell was allowed to communicate with all other cells (Fig. \ref{figures-1}d). We varied the strength of the intercellular communication, which is parameterized by $a_{us}$, and found that it can serve as a control parameter, similar to $a_u$. Starting from low values of $a_{us}$ for which the mlp state is the only stable state, a sequence of bifurcations leads to the appearance and disappearance of several additional stable states. The first stable states to appear have one cell with high levels of $u$ (low levels of $v$) and 3 cells with high levels of $v$ (low levels of $u$). With increasing values of $a_{us}$ the balance shifts to more cells with high $u$. At the extreme there is an interval of $a_{us}$ with only two stable states in which 3 cells have a high level of $u$ and one cell has a high level of $v$. Interestingly, the differentiated states are never homogeneous for the set of parameters used here. \\
Instead of modulating the strength of cell communication by tuning $a_{us}$, the system can also be driven out of the mlp state by increasing cell number, which would happen naturally through cell division. Without cell communication the system remains in the mlp state, irrespective of cell number (Fig. \ref{figures-1}). With cell communication, a symmetry-breaking event occurs and two new stable states appear starting from 4 communicating cells, for the particular parameter set used (Fig. \ref{figures-1}f). For 1 or 2 cells, only the mlp state is stable, and there are no other steady states. From 4 cells on, this state becomes unstable. As demonstrated previously by Stanoev et al. \cite{Stanoev2021} differentiation can occur simply after a certain number of cell divisions without changing the topology of the GRN or imposing a change of its parameters by external cues. Interestingly, more steady states appear in the system with increasing cell numbers. In the following, we will use simulations based on the 4-cell configuration with communication between all cells as ground truth training data for GRN inference with NNs.

\subsection{Feedforward NN regression is unsuitable for GRN parameter inference}
\begin{figure}[!h] 
    \centering
    \includegraphics[width=1\textwidth]{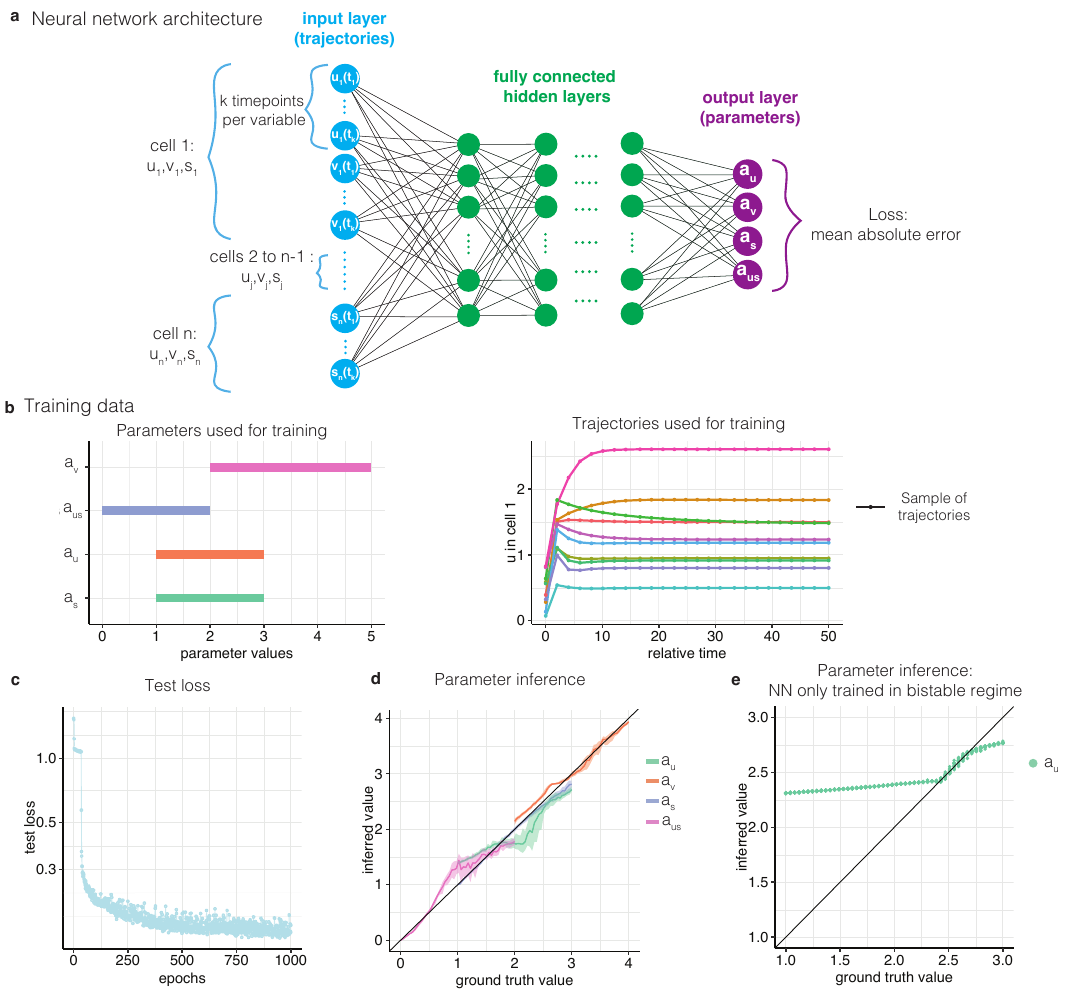}
    \caption{{\bf Regression with a feedforward NN is unsuitable for GRN parameter inference.} a: Architecture of the feedforward NN. The mean absolute error was used for optimization. b: Training data. Left: Parameter ranges used for creating simulated trajectories. Right: 10 example trajectories. c: Test loss during training of the NN. d,e: Ground truth parameter values (used for simulating the trajectories) versus parameter values inferred by the NN. In d, training trajectories covered the mlp as well as the bistable, differentiated regime. In e, the training trajectories came exclusively from the bistable regime.}
    \label{figures-2}
\end{figure}

NNs have shown impressive performance in a large variety of supervised learning tasks \cite{Glorot}. The power of NNs usually relies on the existence of a large amount of high quality training data. Our first, naive idea was therefore to simulate expression trajectories, based on the dynamical system discussed above (see Fig. \ref{figures-1}a), with randomly sampled parameters and use these trajectories to train a feedforward NN regression model (Fig. \ref{figures-2}a). The input layer of this NN consists of the trajectories of $u$,$v$ and $s$ for $n$ cells and $k$ time points. Training samples are therefore vectors of length $3 \cdot n \cdot k$. The output nodes correspond to the 4 parameters of the GRN, $a_u$, $a_v$, $a_s$ and $a_{us}$. Input and output layer were connected by several, fully-connected hidden layers. \\
To test this approach we used a configuration of 4 cells, with communication between all cells (as in Fig. \ref{figures-1}d), and simulated 1000 trajectories with $25$ time points for all variables in all cells. Parameters were sampled uniformly from intervals chosen such that trajectories from both the mlp as well as the differentiated regime were created (Fig. \ref{figures-2}b). Initial states were also chosen randomly within reasonable intervals (see Methods). With this setup, the NN seemed to converge quickly and training was stopped after 1000 epochs (Fig. \ref{figures-2}c). To create the test data we simulated 50 sets of 20 trajectories where the parameters were identical for each trajectory in a set, but the initial states were chosen randomly. Comparison of the parameter values used to simulate the trajectories (ground truth values) with the parameter values inferred by the NN model (Fig. \ref{figures-2}d) revealed good accuracy of the model. Large systematic biases were absent for most parameter values. The random initial conditions contributed to the observed spread around the true values, which might limit the precision of the model.\\
At first glance, the simple feedforward architecture seemed to perform well. We next wanted to test, how important it is that the training data covers the different regimes of the dynamical system. When we trained the model with trajectories from the bistable, differentiated regime we observed that the model performed poorly for test trajectories outside of that regime (Fig. \ref{figures-2}e). As the model is agnostic to the differential equations governing the dynamical system, it was unable to extrapolate beyond the parameter ranges it was trained on. In other words, if trained on a particular regime of the dynamical system, the NN model learns the behavior of that regime and does not generalize well. It would therefore be crucial to cover a large enough area of parameter space with the training data. Importantly, we were only able to identify the correct parameter ranges, because the system is relatively simple, allowing us to obtain a detailed understanding of its qualitative behavior (see Fig. \ref{figures-1}). In an experimental setup, the relevant  parameter ranges are usually unknown and it is typically hard to tune individual parameters. The naive feedforward NN regression model is therefore unsuitable for inferring GRN parameters from experimental data. 

\subsection{Physics informed neural networks can infer GRN parameters from partial and noisy data.}

\begin{figure}[!h] 
\centering
\includegraphics[width=1\textwidth]{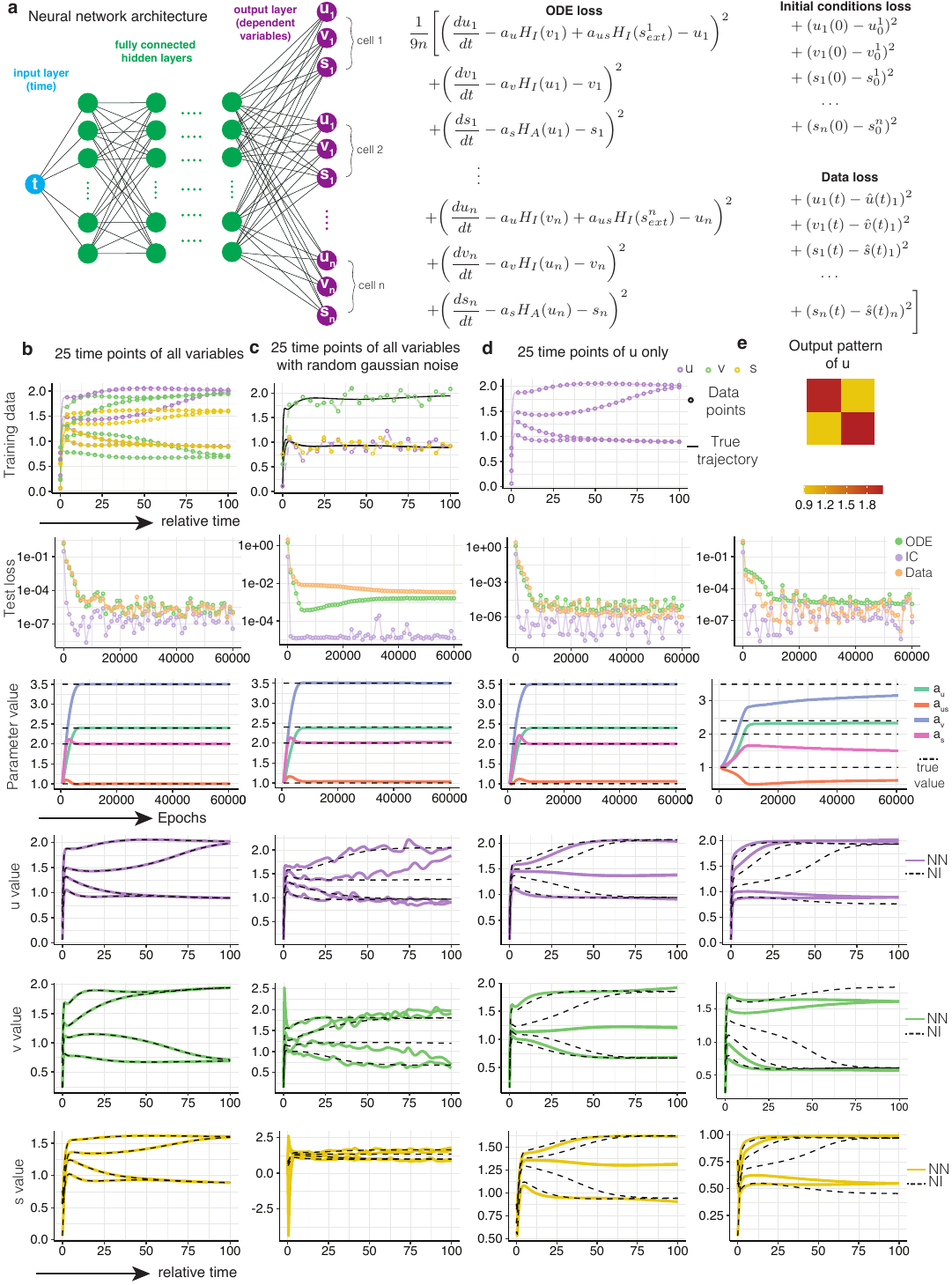}
\caption{{\bf PINN for inference of GRNs from time series data.} See next page for caption.}
\label{figures-3}
\end{figure}

To ameliorate the reliance of NNs on large mounts of training data that represent all regimes of the dynamical system, we have to constrain the inference problem in a meaningful way. Ideally, the NN model should be aware of and respect the underlying differential equations. Physics-informed NNs (PINNs) leverage automated differentiation to solve a broad class of differential equations \cite{Raissi2019, Karniadakis2021}. The input layer of a PINN is composed of the independent variables of the differential equations (such as, for example, space and time for many applications in physics). The PINN is then trained such that the output layer approximates a solution to the differential equations for arbitrary points (in space and time) given as input. Fulfillment of the (ordinary) differential equations as well as initial and boundary conditions is ensured by appropriate loss terms, called ODE, IC and BC loss, respectively. During training, residual points on which the loss terms are evaluated are chosen randomly or in a way that adapts to the particular differential equations \cite{Lu2019}. For this 'forward problem' of finding a solution of fully determined differential equations, no training data is necessary. PINNs can also infer undetermined parameters of the differential equations ('inverse problem'), which does require measured or simulated training data and a corresponding loss term ('data loss') that penalizes deviation of the solution from that data. The loss terms that ensure fulfillment of the differential equations strongly constrain the output space of the NN and thereby reduce the variance of the parameter inference. The issue of poor generalizability we observed with feedforward NN regression (Fig. \ref{figures-2}e) should therefore be absent in PINNs.    \\
To explore whether PINNs can successfully infer GRN parameters ('inverse problem'), we implemented the architecture shown in Fig. \ref{figures-3}a with the DeepXDE package \cite{Lu2019}. The input layer of the NN consists of only one node, which corresponds to time, and the output layer contains all dependent variables ($u$,$v$ and $s$ in all cells). As above, we used the 4-cell configuration with communication between all cells as a proof-of-concept. To generate training data we simulated trajectories with identical parameters but randomly drawn initial states. To explore the limitations of the PINN, we added noise and/or subset the data (Fig. \ref{figures-3}b-e). Starting from noise-free trajectories with $25$ time points per variable (Fig. \ref{figures-3}b), we added Gaussian noise, since measurements are likely noisy due to biological and technical variability (Fig. \ref{figures-3}c). We also explored training the PINN with a subset of variables as it is typically difficult to obtain measurements of all relevant dependent variables in experiments (Fig. \ref{figures-3}d). Lastly, we studied training the model on the first and last time points only. For the set of parameters used here, the system has closely approached a stable steady state with two u-high and two v-high cells (Fig. \ref{figures-3}e, top) by the last time point. 
\addtocounter{figure}{-1}
\begin{figure}[!h] 
\centering
\caption{{\bf PINN for inference of GRNs from time series data.} a: Architecture of the PINN. The input to the network is time and the output consists of all dependent variables of the dynamical system. The PINN is optimized via a loss function that considers the differential equations (ODE loss), the initial conditions (IC loss) and training data (data loss). b, c, d, e: The \textit{first row} shows examples of training scenarios. A GRN with 4 cells that all communicate with each other was used. \textbf{b} Training on noise-free trajectories of all dependent variables with 25 fixed time points. c: Training on trajectories shown on the left with added Gaussian noise. Only the trajectories in one cell are shown. d: Training on noise-free trajectories of $u$ only. e: Only the first and last time point of the $u$ trajectories in all 4 cells were used for training. The \textit{second row} shows the resulting test losses. Colours indicate the different loss terms. \textit{Row 4} shows the inferred parameters and \textit{rows 3 - 6} show the approximated trajectories for the four scenarios. In the trajectory plots solid lines are trajectories approximated by the PINN and dashed lines are trajectories calculated by numerical integration using the inferred parameters.}
\end{figure}
This scenario is relevant for measurements with only one or a few time points or if the system is practically always in a stable steady state. In Fig. \ref{figures-3}b-e we give examples of model behavior for different training scenarios. A systematic exploration and quantification of model performance is presented in Fig. \ref{figures-4}.\\
When using complete trajectories for training, the PINN converges robustly after a few epochs and all three loss terms have similar convergence rates (Fig. \ref{figures-3}b, second row). The inferred parameters are close to the ground truth parameters (Fig. \ref{figures-3}b, third row) and the trajectories approximated by the PINN coincide with the trajectories calculated by numerical integration using the inferred parameters (Fig. \ref{figures-3}b, rows 4-6). In contrast to feedforward NN regression (Fig. \ref{figures-2}), which required many training samples, the PINN needs only one set of trajectories for accurate GRN inference. As to be expected, noise reduced the performance of the model, likely due to over-fitting, which can be seen for the inferred trajectories in Fig. \ref{figures-3}c. Model performance was also compromised when only one dependent variable was used for training (Fig. \ref{figures-3}d). Providing only the initial and final time point presented the biggest challenge for the PINN (Fig. \ref{figures-3}e): The trajectories approximated by the PINN show large discrepancies with the trajectories calculated by numerical integration using the same (inferred) parameters. Hence, the trajectories approximated by the PINN are not a proper solution of the differential equations. Surprisingly, the inferred parameters were still roughly correct.\\
For a more systematic and quantitative assessment of model performance we tested $84$ different conditions and considered: 1. the mean squared error between trajectories approximated by the PINN and trajectories found through the numerical integration using the inferred parameters , 2. the test loss and 3. the relative error of the inferred parameters (Fig. \ref{figures-4}). For each condition we averaged over $10$ runs with identical GRN parameters but randomly drawn initial states. First, we focused on the training scenarios that utilized all time points (Fig. \ref{figures-4}b-d). As to be expected, increasing levels of noise reduced model performance (Fig. \ref{figures-4}b). 
\begin{figure}[!h] 
\centering
\includegraphics[width=1\textwidth]{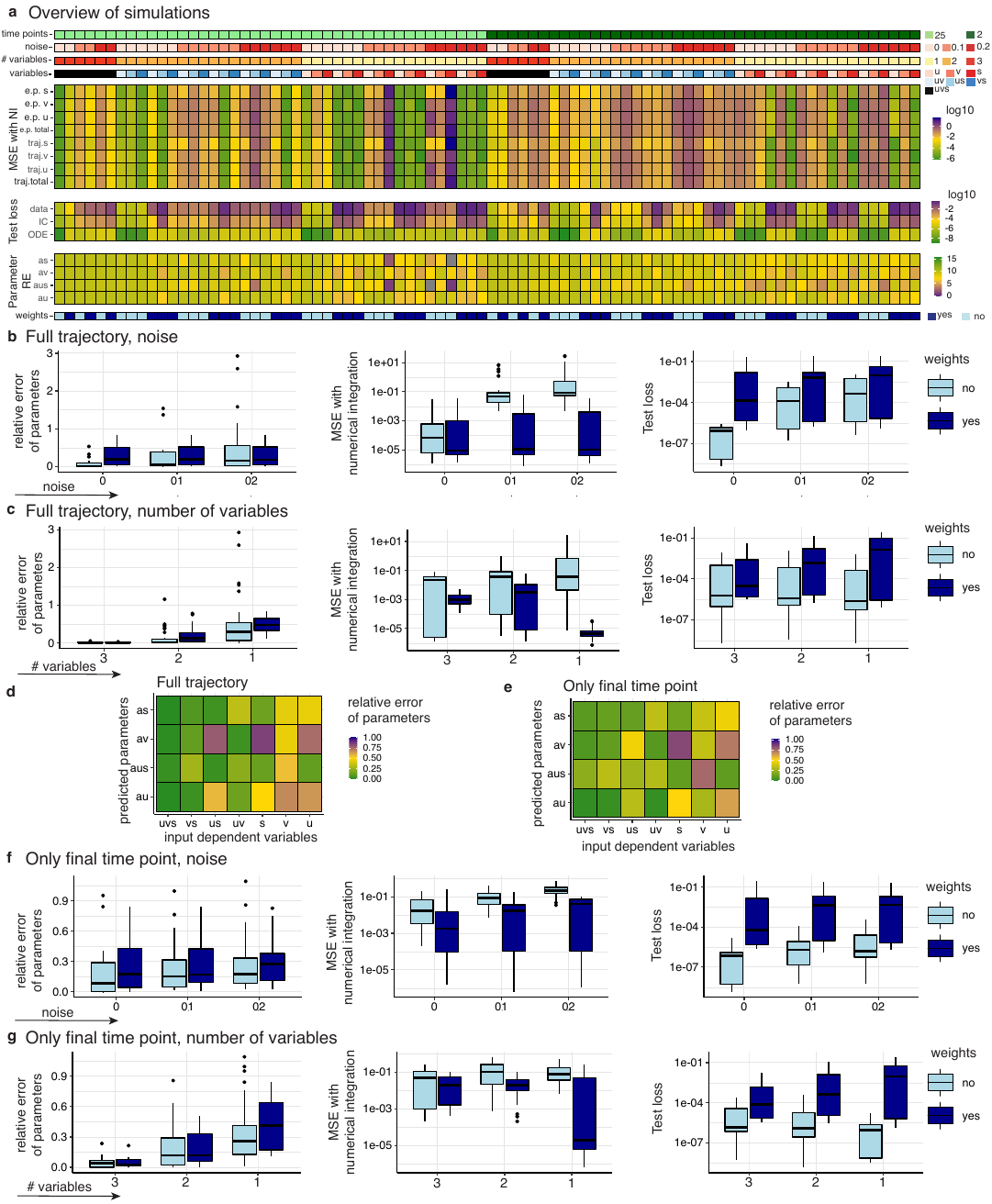}
\caption{{\bf PINNs can infer GRN parameters from partial and noisy time series data.}  a: Overview of all simulation conditions. The following parameters of the training data set were varied: the number of time points (either 25 per variable or only the first and last time point), the amount of Gaussian noise (standard deviation of $0$, $0.1$ or $0.2$ with a mean of $0$) and the dependent variables ([u,v,s], [u,v], [u,s], [v,s], [u], [v], [s]). When weights were given to the loss terms the ODE loss was weighted with a factor 1000. e.p.: end point, traj: full trajectory. b-d: Dependence of PINN performance on noise level (b), number (c) and identity (d) of dependent variables used for training when the complete trajectories were used. e-g: Same performance comparisons as in b-d but the PINN was trained only on the initial and final time point of the trajectories.}
\label{figures-4}
\end{figure}
In an attempt to mitigate over-fitting to the noisy training data, we introduced weights for the three loss terms and gave the ODE loss a 1000-times higher weight. Weighting improved trajectory approximation, but did not have a strong influence on parameter inference. Removing dependent variables from the training set had a strong and systematic effect on parameter inference and trajectory approximation was similarly affected when no weights were used (Fig. \ref{figures-4}c). Weighting strongly improved trajectory approximation when only one dependent variable was used for training. Importantly, the relative errors of the parameter values depended on the set of variables used for training (Fig. \ref{figures-4}d). For example, when only $u$ and $v$ were used, the parameters $a_u$ and $a_v$ were inferred more accurately than the parameters $a_{us}$ and $a_s$. Conversely, when only $u$ and $s$ were used, $a_{us}$ and $a_s$ had a smaller error than the other parameters. Learning from only the first and last time point of the training data was overall a harder task for the PINN (Fig. \ref{figures-4}e-g), but we observed similar trends for the dependence of model performance on noise (Fig. \ref{figures-4}f) or the number of dependent variables used for training (Fig. \ref{figures-4}g). Surprisingly, parameter inference from two time points was almost as accurate as when the whole trajectories were used for training, while the PINN's approximation of the trajectories was compromised. In summary, the PINN was able to infer GRN parameters even when only partial data was supplied for training.
\subsection{PINN for inference of GRNs from snapshot data.}
\begin{figure}[!h] 
    \centering
    \includegraphics[width=1\textwidth]{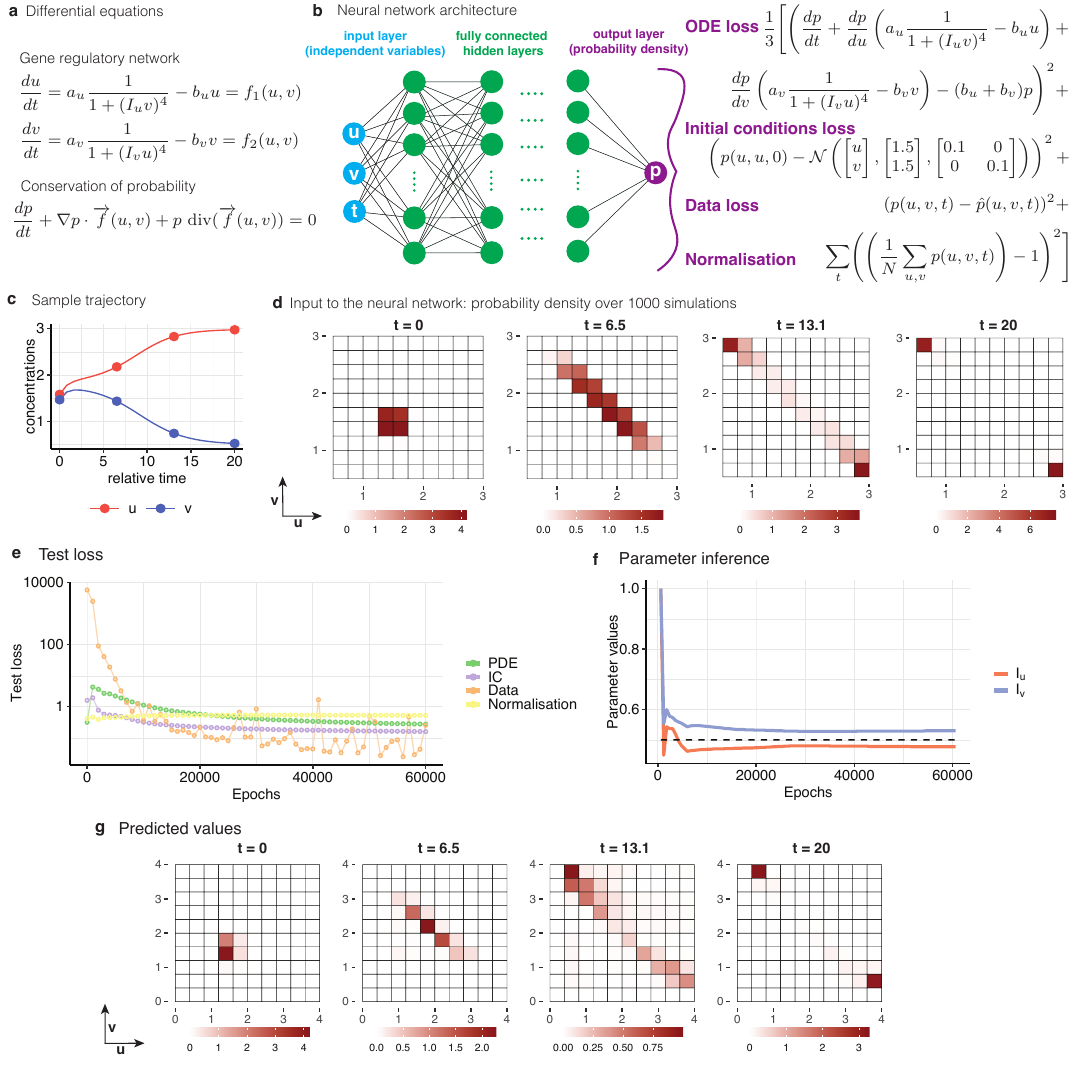}
    \caption{{\bf PINN for inference of GRNs from snapshot data.} a: Top: Differential equations describing two, mutually inhibiting genes $u$ and $v$. Bottom: Differential equation describing the conservation of probability. b: Architecture of the PINN. The input nodes correspond to $u$, $v$ and time $t$. The output is the probability density. The PINN is optimized via a loss function that considers the differential equations, the initial conditions, the training data and the normalisation. c: Example trajectory for the differential equations shown in panel a. d: Probability density of the values $u$ and $v$ at four different time points, generated by repeated simulation of trajectories as in c. This simulation was used as training data. e: Test loss of PINN training using the data shown in d. f: Convergence of the parameters $I_u$ and $I_v$ during PINN training. g: Probability densities approximated by the PINN at $4$ different time points.}
    \label{figures-5}
\end{figure}

Most high-throughput single-cell profiling assays are destructive, which prevents the measurement of single-cell trajectories. These assays therefore only provide "snapshots" of the system dynamics. Additionally, in conventional single-cell omics experiments, any information about the spatial arrangement of the cells is lost. Therefore we wanted to explore, how a PINN would perform when trained with snapshot data that lacks spatial resolution. As any parameter related to cell communication is unlikely to be estimated well in such a scenario, we considered a simpler dynamical system of two mutually inhibiting genes, $u$ and $v$, without cell communication \cite{Ferrell2012} (Fig. \ref{figures-5}a). The parameters $I_u$ and $I_v$ modulate the inhibition of $u$ or $v$, respectively, by the other gene. For a particular set of parameters, the dynamical system is bistable and leads to the cell autonomous differentiation into either a u-high or a v-high state. Cells will be attracted to one of these stable steady states depending on their initial state. As trajectories cannot be obtained from destructive snapshot measurements it is more natural to model the system at the level of a population of cells and consider a bivariate probability density of $u$ and $v$. In the absence of noise, the dynamics of the probability density is completely determined by the conservation of probability (Fig. \ref{figures-5}a), which is therefore the differential equation that must be fulfilled by the PINN.\\
Fig. \ref{figures-5}b shows the architecture of the PINN together with the loss terms. The input layer is now composed of three nodes, corresponding to $u$, $v$ and time $t$. The only output node is the probability density at the point [$u$,$v$,$t$] given as input. As before, the loss considers the governing differential equations, initial conditions and the training data. For training we simulated 1000 trajectories with initial values of $u$ and $v$ randomly drawn from a bivariate normal distribution centered around $1.5$. As mentioned above, the parameters of the GRN and the distribution of the initial states are chosen such that trajectories tend to one of two stable steady states (Fig. \ref{figures-5}c). The simulated trajectory positions were binned for each time point and the probability densities were approximated by the relative frequencies (Fig. \ref{figures-5}d). As intended, the initial probability density, a normal distribution, developed  into a bimodal distribution, reflecting the existence of two stable steady states with anti-correlated expression of $u$ and $v$. Using these simulations we trained the PINN, leaving the parameters $I_u$ and $I_v$ undetermined. 
The PINN converged quickly (Fig. \ref{figures-5}e) and the parameters were inferred with reasonable  precision (Fig. \ref{figures-5}f). The probability densities approximated by the PINN show qualitatively the same dynamics as the densities used for training (Fig. \ref{figures-5}g). However, the PINN approximation of the probability density had a few negative values and was not always properly normalized, which could likely be improved by additional constraints. 

\begin{figure}[!h] 
    \centering
    \includegraphics[width=1\textwidth]{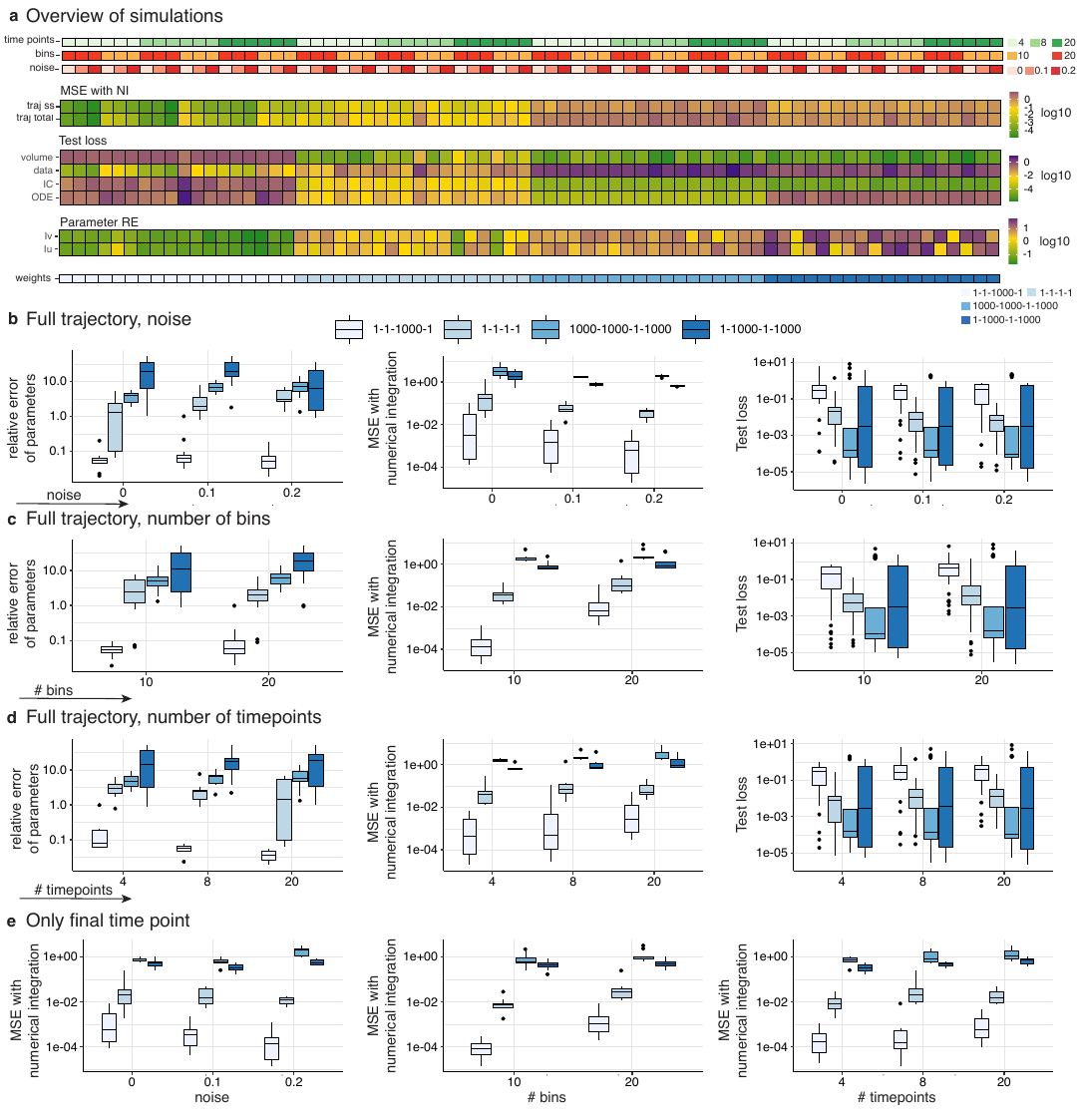}
    \caption{{\bf PINNs can infer GRN parameters from snapshot data for several training time points and noise levels.} a: Overview of all simulation conditions. The following parameters of the training data set were varied: the number of time points (either 4, 6 or 20 per variable), the amount of Gaussian noise (standard deviation of $0$, $0.1$ or $0.2$ with a mean of $0$) and the number of bins for $u$ and $v$ (either 10 or 20 per variable). Different weights of the loss function were tested during training. The loss function was composed of: PDE loss, IC loss, data loss and normalisation loss. The weights are given in this sequence. traj ss: trajectory at steady state, traj total: full trajectory. b-d: Dependence of PINN performance on noise level (b), number of bins (c) and number of time points (d) of dependent variables used for training when the complete trajectories was evaluated. e: Same performance comparisons as in b-d but the PINN was evaluated only on the steady state of the system.}
    \label{figures-5-2}
\end{figure}

For a more systematic assessment of model performance we tested $75$ different conditions and considered, as above: 1. the mean squared error between trajectories approximated by the PINN and trajectories found through the numerical integration using the inferred parameters , 2. the test loss and 3. the relative error of the inferred parameters (Fig. \ref{figures-5-2}). For each condition we averaged over $5$ runs with identical GRN parameters but randomly drawn initial states. Examples for prediction results can be found in \nameref{S1_Fig}. First, we focused on the scenarios that utilize the entire predicted trajectory for comparison (Fig. \ref{figures-5-2}b-d). As expected, increasing levels of noise reduced model performance (Fig. \ref{figures-5-2}b) but a strong weight on the data loss seemed to mitigate that effect. The number of bins used to aggregate individual trajectories only had an influence on the prediction of the probability density (Fig. \ref{figures-5-2}c). Surprisingly, fewer bins were advantageous, which might be due to averaging over noise related to the relatively small amount of simulated trajectories. For parameter inference, using more time points seemed beneficial, in particular when a high weight was placed on the data loss (Fig. \ref{figures-5-2}d). Finally, when only the last time point was used for testing, we observed that the final, steady state was predicted with good accuracy (Fig. \ref{figures-5-2}e). Across all considered scenarios, a high weight on the data loss was typically advantageous, which highlights the importance of high quality training data even in our highly constrained approach. All in all, the PINN was able to infer GRN parameters from snapshot data for a model without cell communication. 
\section{Discussion}
The inference of GRNs from noisy and usually incomplete measurements is a long-standing challenge which inspired the development of many different approaches. In this manuscript, we studied the performance of PINNs in this context. \\
PINNs are general tools that approximate the solutions to a broad class of differential equations. To apply them to GRN inference requires expressing the dynamical system defined by the GRN as a set of differential equations. To that end, specific expressions that model the gene interactions have to be assumed. Here, we used Hill functions with fixed Hill coefficients for both activation and inhibition. We selected a subset of relevant parameters to be learned by the PINN, but it might be interesting to leave more parameters undetermined, especially the Hill coefficients. Most importantly, we used the same network topology for simulation and training the PINN: The same genes were connected with the same type of interaction (either activating or inhibiting). In principle, one could base the training on a fully connected network and model each interaction as the sum of activating and inhibiting expressions. Such a setup would leave network topology unconstrained and it would be interesting to explore, if a PINN could infer it from the data. In this context, it might be useful to add a regularization term to the loss function such that only the strongest interactions are selected and the inferred GRN is sparse.\\
As a proof of concept we studied a minimal GRN with two mutually inhibiting genes. Such a GRN exhibits a bifurcation that models the differentiation of a multipotent progenitor into one of several differentiated cell types. While several pairs of master transcription factors that govern such bifurcations have been identified in experiments, real GRNs contain other relevant genes. It would therefore be useful to determine, how much and what kind of experimental data would be necessary to infer a much larger GRN with a PINN. \\
In this study we first considered an experimental scenario in which the trajectories of individual cells are measured and demonstrated that a PINN outperforms a simple feedforward NN regression model. While the feedforward NN requires many training samples that must cover all dynamical regimes of the GRN, the PINN efficiently infers GRN parameters from a single sample, at the cost of making assumptions or using prior knowledge about the GRN. The PINN was able to infer parameters even if only a subset of dependent variables was used for training. The relative errors of the inferred parameter values depended on the identity of the used variables, which is important to keep in mind for optimal experiment design. As the training of the PINN becomes computationally more costly with increasing number of cells, it would be interesting to explore, whether using measurements of a subset of cells for PINN training is sufficient for GRN inference. Possibly, that would require a kind of mean field approximation of the cells that are not used for training. Surprisingly, parameter inference was still possible when we only used the initial and final time point of the trajectories for training. However, in this case, the approximate trajectories provided by the PINN did not fulfill the differential equations as they deviated from trajectories calculated by numerical integration using the inferred parameters. It seems that the loss terms related to additional time points support the ODE loss in ensuring fulfillment of the differential equation. Inferring the GRN from the final state, which is in this case essentially a spatial pattern of differentiated cell types, would be very useful not only to study morphogenesis but also to inform synthetic biology applications. Recently, NNs were used to implement a cellular automaton that models morphogenesis \cite{Mordvintsev2020}. Impressively, it was shown that providing a desired spatial pattern as training data is sufficient to train the NN such that the automaton robustly develops into that spatial pattern. While this is certainly a conceptually important feat, cellular automata are only rough approximations of real biological dynamics and it would be preferable to achieve a similar performance with GRNs. We speculate that constraining the final state to be a globally stable steady state, potentially by using a Lyapunov function \cite{Haustenne2015}, might help in that respect. \\
In the final section of this manuscript we studied a scenario in which only snapshot data of cell populations are available, which is the case for single-cell RNA-sequencing experiments \cite{Kolodziejczyk2015}. As spatial context is not available in this scenario, we described the system at the population level, with probability densities of gene abundances. We showed that a PINN was able to infer a simple GRN without cell communication. While parameter inference was successful, the probability density approximated by the PINN sometimes deviated from proper normalisation as we used a soft constraint implemented by the normalisation loss.  To enforce proper normalization one could reformulate the differential equations such that the dependent variable is a normalized function \cite{Uy2020}. Another option is to discretize the domains of $u$ and $v$ and add a constraint on the sum of the resulting discrete densities via an additional loss term \cite{Xu2020NN}. As there was no noise in the dynamical system used here, the relevant differential equation was simply given by the conservation of probability. In the presence of noise, one would have to use the Fokker-Planck equation whose parameters should in principle be inferrable by a PINN, as well.\\ 
Next to the two experimental scenarios considered in this study, there are a few others that are currently very popular and would therefore be worthwhile to explore in future work. Many snapshot measurements of highly dynamical systems, such as developing tissues, in fact contain dynamical information: Pseudotime methods have been used to establish developmental progression from snapshot data \cite{Haghverdi2016, Ji2016}. It would be interesting to work out, how pseudotime information could be leveraged for GRN inference with PINNs. Spatially resolved omics modalities are also being used extensively at the moment. To infer GRNs with cell communication from such data will be an interesting challenge. \\
\section*{Conclusion}
In conclusion, we have established that PINNs can be used for the accurate inference of GRNs. PINNs thus present an exciting, new way to obtain mechanistic insights from single-cell data. We hope that our work will stimulate colleagues from mathematics, physics and biology to collaborate on the many fascinating problems presented by single-cell developmental biology.\\
\section*{Supporting information}

\paragraph*{S1 Fig.}
{\bf Overview of PINN predictions for snapshot data.} a-g: Predicted probability densities for different training data sets and hyper-parameters. (a) shows the parameter combination for the best prediction. b-g show results where one parameter has been changed from (a): (b) number of bins is 20, (c) noise level with standard deviation 0.2, (d) weights are set to 1-1-1-1, (e) weights are set to 1000-1000-1-1000, (f) weights are set to 1-1000-1-1000, (g) 20 time points were used for training. (h) shows sampled trajectories without noise, Gaussian noise with 0.1 standard deviation and Gaussian noise with 0.2 standard deviation. Solid lines are the true trajectories and points are the training data points used.
\label{S1_Fig}

\section*{Acknowledgments}
The authors acknowledge funding by the Netherlands Organisation for Scientific Research (NWO/OCW \url{www.nwo.nl}), as part of the Frontiers of Nanoscience (NanoFront) program. D. G. acknowledges support from SoBigData.it (\url{https://pnrr.sobigdata.it/}), which receives funding from European Union - NextGenerationEU - National Recovery and Resilience Plan (Piano Nazionale di Ripresa e Resilienza, PNRR), project `SoBigData.it - Strengthening the Italian RI for Social Mining and Big Data Analytics' - Grant IR0000013 (n. 3264, 28/12/2021).
The computational work was carried out on the Dutch national e-infrastructure with the support of SURF Cooperative.


\bibliography{PINN_bib}

\newgeometry{left=15mm,top=15mm}
\begin{figure}[!h] 
    \centering
    \includegraphics[width=1\textwidth]{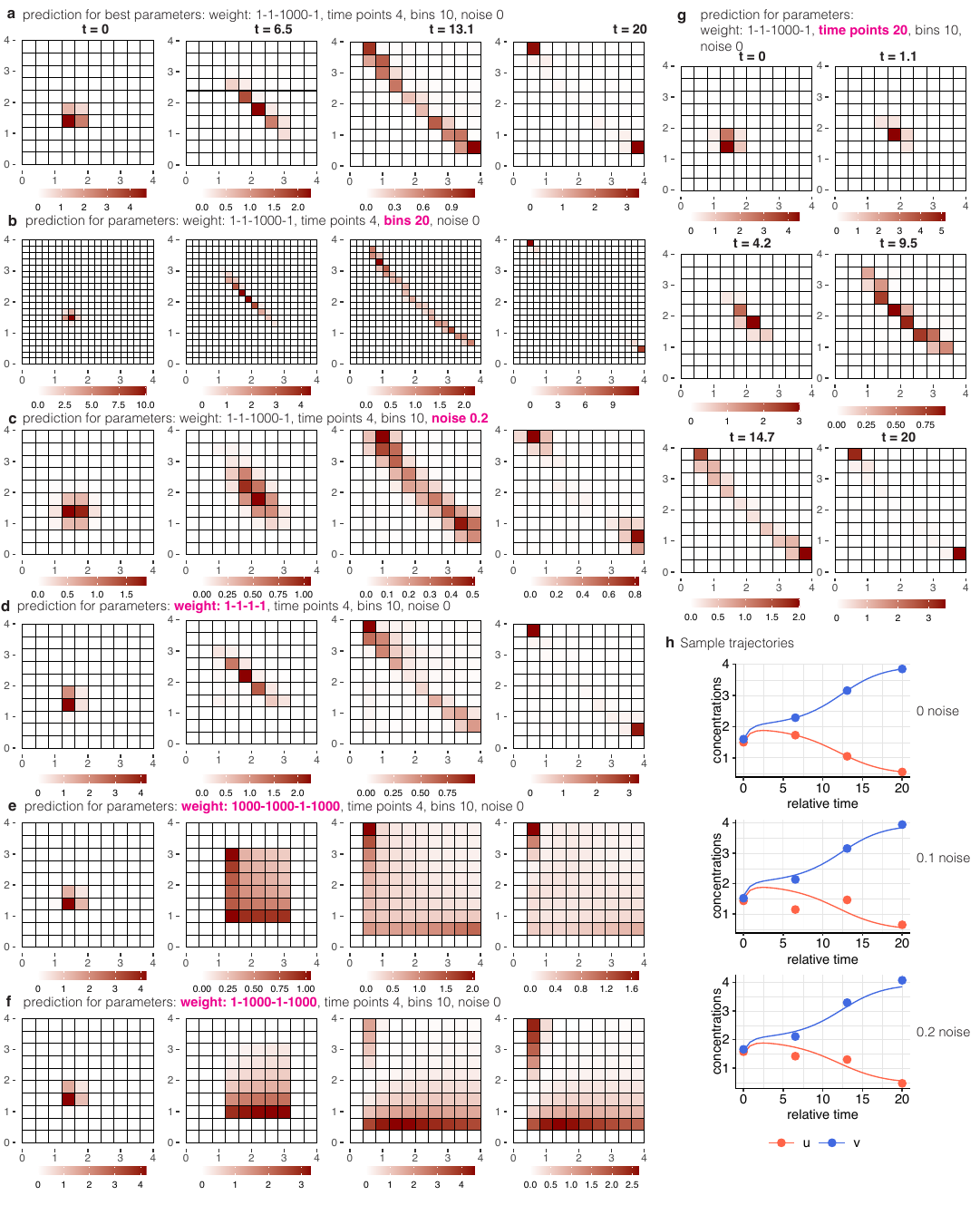}
    \caption*{{\bf Figure S1}}
\end{figure}
\restoregeometry  

\end{document}